# The Evolving Baryonic Tully–Fisher Relation: A Universal Law from Galaxies to Clusters

Stuart Marongwe [1*], Stuart A. Kauffman[2]
1. Physics Department, University of Botswana 4775 Notwane Road, Gaborone, Botswana, email: stuartmarongwe@gmail.com

2. University of Pennsylvania 34th & Spruce Street, Philadelphia, PA, 19104-6303 (emeritus) stukauffman@gmail.com

* corresponding author

## Abstract
The baryonic Tully–Fisher relation (BTFR) links the baryonic mass of galaxies to their characteristic rotation velocity and has been shown to hold with remarkable precision across a wide mass range. Recent studies, however, indicate that galaxy clusters occupy a parallel but offset relation, raising questions about the universality of the BTFR. Here, we demonstrate that the offset between galaxies and clusters arises naturally from cosmic time evolution. Using the evolving BTFR derived from the Nexus Paradigm of quantum gravity, we show that the normalization of the relation evolves as an exponential function of cosmic time, while the slope remains fixed at ~4. This provides a simple and predictive framework in which both galaxies and clusters obey the same universal scaling law, with their apparent offset reflecting their different formation epochs. Our results unify mass–velocity scaling across five orders of magnitude in baryonic mass, offering new insights into cosmic structure formation.

## 1. Introduction

Scaling relations serve as fundamental pillars in extragalactic astrophysics, unveiling underlying patterns amid the intricate processes governing the formation and evolution of galaxies and galaxy clusters. Among these, the Tully–Fisher relation (TFR; Tully & Fisher 1977) stands out as a key empirical connection between a galaxy's luminosity and its rotational velocity. Its baryonic extension, the baryonic Tully–Fisher relation (BTFR; McGaugh et al. 2000; Lelli, McGaugh & Schombert 2016), refines this by linking the total baryonic mass (stars plus gas) to the characteristic rotation velocity, exhibiting a remarkably tight correlation with a slope near 4 and intrinsic scatter as low as 0.1 dex across diverse disk galaxies. Extending this relation to the most massive gravitationally bound structures in the universe such as galaxy clusters, presents a compelling challenge. Unlike galaxies, clusters do not exhibit coherent rotation, yet analogous velocity measures, such as galaxy velocity dispersions or circular velocities derived from X-ray observations and gravitational lensing, enable meaningful comparisons. Early theoretical explorations (Sanders 1994; Mo, Mao & White 1998) hinted at a potential continuity, but contemporary analyses (Gonzalez et al. 2013; Sadu et al 2024; Chiu et al. 2018) reveal that clusters trace a parallel BTFR as depicted in Figure 1, albeit offset from the galactic relation by approximately 0.6–0.8 dex in logarithmic baryonic mass. This discrepancy prompts a critical inquiry: do galaxies and clusters adhere to separate scaling laws, shaped by distinct physical mechanisms, or does a unified, time-evolving framework reconcile them? In this study, we affirm the latter perspective. We apply the evolving BTFR framework proposed by Marongwe (2024), which introduces a cosmic time-dependent normalization while preserving a constant slope of ~4, we illustrate that the observed offset arises naturally from differences in formation epochs—galaxies typically assembling at higher redshifts (z ~ 2–3) compared to clusters (z < 1). This approach not only bridges the gap between these cosmic structures but also extends the BTFR's applicability across five orders of magnitude in baryonic mass, yielding profound implications for our understanding of hierarchical structure formation in the Lambda Cold Dark Matter paradigm.

## 2. The Baryonic Tully-Fisher Relation: an overview

### 2.1 The Galactic BTFR

The baryonic Tully–Fisher relation (BTFR) represents a cornerstone empirical scaling law in galactic astrophysics, linking the total baryonic mass $M_b$ (encompassing stars and gas) of rotationally supported galaxies to their characteristic rotation velocity $v$, typically expressed as $\log M_b = n \log v + c$, where $n \approx 4$ and $c$ is the normalization constant. Originally derived from the classical Tully–Fisher relation (TFR), which correlates luminosity with rotational width (Tully & Fisher 1977), the BTFR was formalized by McGaugh et al. (2000) to incorporate baryonic content, addressing limitations in stellar-mass-only formulations. This extension has proven particularly robust for disk galaxies with well-measured flat rotation curves extending beyond the baryonic scale radius, exhibiting a slope of 3.8–4.0 and remarkably low intrinsic scatter of ~0.1 dex (Lelli et al. 2019).

Subsequent studies have expanded the BTFR's applicability across a broad dynamic range. For instance, the Spitzer Photometry and Accurate Rotation Curves (SPARC) sample has been instrumental in refining the relation, confirming its tightness and universality for late-type spirals (Lelli et al. 2019; McGaugh 2012). Recent large-scale surveys, such as Cosmicflows-4, have extended this to approximately 10,000 galaxies, providing distance-independent calibrations and highlighting the BTFR's role in constraining cosmological parameters like the Hubble constant (Kourkchi et al. 2022).

At the low-mass end, investigations into dwarf and ultra-diffuse galaxies (UDGs) have tested deviations, with gas-rich UDGs aligning well on the BTFR despite their extended morphologies (Mancera Piña et al. 2024). Ultrafaint dwarfs, however, pose challenges, due to possibly non-equilibrium dynamics that could imply offsets, though recent analyses suggest consistency when accounting for measurement uncertainties (Müller 2019).
Near-infrared photometry from WISE and Spitzer has further bolstered stellar mass estimates, enhancing the BTFR's precision for kinematic studies (Schombert et al. 2024).
Theoretical interpretations often invoke Modified Newtonian Dynamics (MOND) or dark matter halo models to explain the relation's origin, with the slope near 4 aligning with MOND predictions (McGaugh 2005).

### 2.2 Extension of the BTFR to Galaxy Clusters

While the BTFR is well-established for galaxies, its extension to galaxy clusters requires adapting velocity proxies due to the absence of coherent rotation. Common alternatives include galaxy velocity dispersion, circular velocities derived from X-ray hydrostatic equilibrium profiles, and mass estimates obtained through weak gravitational lensing. These approaches allow for meaningful comparisons between the scaling relations observed in galaxies and those in clusters, despite the differences in their internal kinematics and the methods used to measure their characteristic velocities.
Early explorations suggested a possible continuity (Sanders 1994; Mo, Mao & White 1998), but empirical compilations reveal clusters following a parallel relation with a slope consistent with ~4, yet offset by 0.6–0.8 dex toward higher normalization compared to galaxies (Gonzalez et al. 2013; Chiu et al. 2018). Key studies have quantified this disparity. For example, Gonzalez et al. (2013) analyzed baryon fractions in clusters, finding alignment with a mass-velocity scaling but with elevated baryonic content at fixed velocity. Sadhu et al. (2024) introduced the "baryonic Faber–Jackson relation" (BFJR) for groups and clusters, analogous to the BTFR but using velocity dispersion, confirming a slope of ~4 and low scatter in intermediate-mass systems.
Mistele (2025) provided a comprehensive quantification, incorporating effective radii and luminosities across scales from globular clusters to galaxy clusters, affirming the offset while exploring a unified Fundamental Plane. Recent weak-lensing reconstructions of cluster mass models, however, challenge the offset's universality, suggesting that clusters may align with the galactic BTFR when avoiding hydrostatic bias which could potentially reconcile the relations under certain dynamical assumptions (Mistele et al. 2025). In MOND contexts, extensions to clusters predict similar scalings but require adjustments for environmental effects (Milgrom 2010). These findings highlight ongoing debates about whether the offset reflects distinct physics, such as enhanced baryon retention in clusters, or systematic measurement differences.

## 2.3 Cosmic Evolution

The hierarchical nature of structure formation in the Lambda Cold Dark Matter (ΛCDM) paradigm implies that scaling relations like the BTFR may evolve with cosmic time, as galaxies assemble at higher redshifts (z ~ 2–3) compared to clusters (z < 1). This temporal disparity is pivotal for interpreting offsets, with early models proposing density-dependent evolution (Mo et al. 1998). Observational evidence supports modest redshift dependence: Puech et al. (2010) traced the stellar and baryonic TFR from z = 0.6 to z = 0, finding significant zero-point shifts indicative of mass growth at fixed velocity. High-redshift studies further illuminate this. At 0.6 ≤ z ≤ 2.5, the TFR exhibits a gradually evolving slope and zero-point, with disks at fixed rotation velocity appearing less massive at earlier epochs (Tiley et al. 2018; Übler et al. 2017). Between z ~ 2.3 and z ~ 0.9, the baryonic TFR shows constant slope but an evolving normalization, consistent with gas accretion and star formation histories (Simons et al. 2017; Tiley et al. 2018). Simulations, such as those from the SIMBA suite, replicate this redshift evolution, attributing it to baryonic physics and feedback processes (Glowacki et al. 2021).
Marongwe (2024) advances this by deriving an explicit time-dependent form, where normalization scales exponentially with cosmic time via $e^{-4H_0 t}$, providing a predictive mechanism for offsets between early-forming galaxies and late-assembling clusters. This framework unifies evolutionary trends, offering testable predictions for future surveys like those with the James Webb Space Telescope.

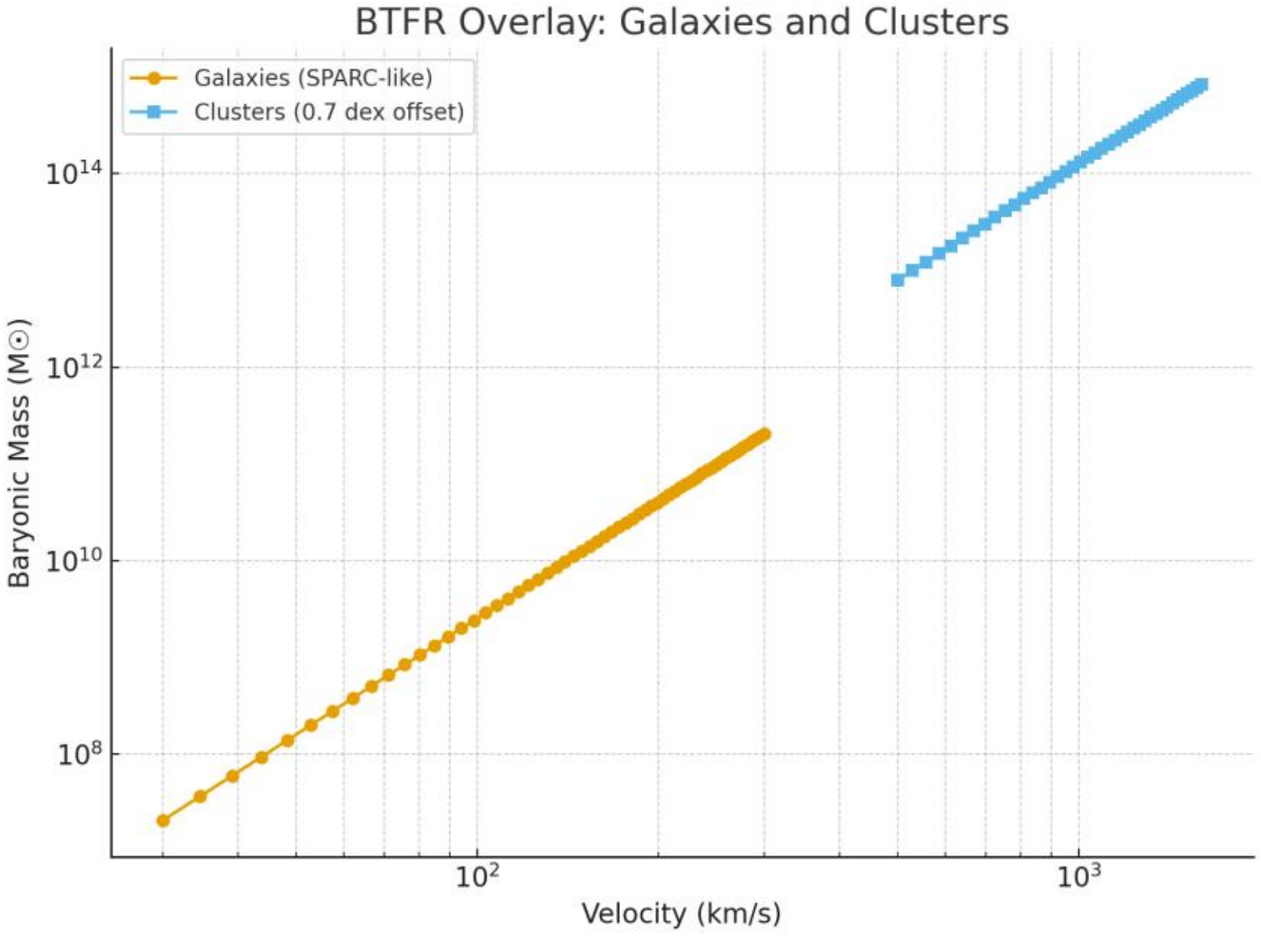

**Figure 1. The Baryonic Tully–Fisher Relation Across Galaxies and Clusters.** The empirical baryonic Tully–Fisher relation (BTFR) constructed from representative literature datasets spanning disk galaxies and galaxy clusters. Galaxy data follow the classical slope-4 relation $M_b \propto v^4$with low intrinsic scatter, as established by SPARC and related surveys. Cluster data—based on velocity dispersions, X-ray–derived circular velocities, and weak-lensing mass profiles—trace a nearly parallel BTFR with an offset of ≈0.6–0.8 dex toward higher baryonic mass at fixed velocity. This offset, long interpreted as evidence for distinct physical regimes, is re-examined in this work and shown to arise naturally from differences in formation epoch under the evolving BTFR predicted by the Nexus Paradigm.

### 3. Methodology

The evolving baryonic Tully–Fisher relation (BTFR) adopted in this study is rooted in the Nexus Paradigm of quantum gravity, as detailed in Marongwe (2024). This theoretical framework posits that spacetime is quantized into discrete Bloch wave packets, akin to phonons in a solid-state lattice, where gravitational phenomena emerge from perturbations in this quantized medium. In this paradigm, dark matter is interpreted as localized vacuum energy in the form of Ricci solitons which are stable, soliton-like solutions to the Ricci flow equations that describe the curvature evolution of spacetime, while dark energy arises from a Higgs-like scalar field with negative energy density. Baryonic matter interacts with this quantized spacetime, inducing scale-invariant accelerations that modify classical Newtonian dynamics at low acceleration regimes, similar to Modified Newtonian Dynamics (MOND; Milgrom 1983). Central to this work is equation (31) from Marongwe (2024), which encapsulates the time-dependent BTFR in the Nexus Paradigm:

$$v \propto e^{H_0 t} M_b^{\frac{1}{4}} \quad , \tag{1}$$

where $v$, is the characteristic rotation velocity (or velocity proxy for clusters), $M_b$ is the baryonic mass, $H_0$ is the Hubble constant in a pure De Sitter phase, which in the Nexus Paradigm is the fundamental frequency of the ground state Ricci soliton of De Sitter (dS) topology and $t$, is the cosmic time elapsed since the structure's formation epoch (lookback time). In the NP, two types of solitons exist: vacuum De Sitter solitons with quantum states $n^2 = \frac{r_H^2}{r_n^2}$ where $r_H$ is the Hubble radius and anti-De Sitter (AdS) solitons with quantum states $\tilde{n}^2 = \frac{r_n c^2}{GM(r)}$ associated with baryonic matter.
Spacetime is therefore described in terms of dS, AdS and a ground state dS soliton for dark energy in the form

$$G_{(nk)\mu v} = (\tilde{n}^2 + n^2 - 1)\Lambda g_{(nk)\mu v} \tag{2}$$

Einstein's field equations are presented here in purely geometric terms, describing a compact Einstein manifold. For any quantum state in which a Ricci soliton exhibits constant curvature, energy remains conserved. The right side of the equation is a symmetric tensor representing the quantum or energy state of spacetime, while the left side functions as a Laplacian operator that averages the trajectories of a test particle within the gravitational field of the given quantum state.

The linearized Eq.(2) is solved by representing it as a Ricci soliton in the $N$-th quantum state, resulting in the equation:

$$G_{(Nk)\mu\upsilon} = N^2 \Lambda g_{(Nk)\mu\upsilon} \quad (3)$$

**3.1 Derivation of the Evolving BTFR**

The exact solution for equation (3) is solved as

$$ds^2 = -\left(1 - \left(\frac{2}{N^2}\right)\right) c^2 dt^2 + \left(1 - \left(\frac{2}{N^2}\right)\right)^{-1} dr^2 + r^2(d\theta^2 + sin^2\theta d\varphi^2)$$

$$= -\left(1 - \left(\frac{2GM_N}{rc^2}\right)\right) c^2 dt^2 + \left(1 - \left(\frac{2GM_N}{rc^2}\right)\right)^{-1} dr^2 + r^2(d\theta^2 + sin^2\theta d\varphi^2) \quad (4)$$

Here $M_N(r) = M_B(r) + M_{DM}(r) + M_\Lambda(r)$ where the terms on right represent the baryonic mass, the DM mass and the DE mass enclosed inside a sphere of radius $r$. This yields a metric equation of the form

$$ds^2 = -\left(1 - 2\left(\frac{GM_b}{rc^2} + \frac{H_0 vr}{c^2} - \frac{H_0 cr}{2\pi c^2}\right)\right) c^2 dt^2 + \left(1 - 2\left(\frac{GM_b}{rc^2} + \frac{H_0 vr}{c^2} - \frac{H_0 cr}{2\pi c^2}\right)\right)^{-1} dr^2 + r^2(d\theta^2 + sin^2\theta d\varphi^2) \quad (5)$$

Where $\frac{GM_{DM}(r)}{r} = v^2 = (H_0 r)^2 = H_0 vr$ and $\frac{GM_\Lambda(r)}{r} = -\frac{H_0 cr}{2\pi}$ The above metric equation leads to the following equation for gravity

$$\frac{d^2r}{dt^2} = \frac{GM_b}{r^2} + H_0 v - \frac{H_0 c}{2\pi} \quad (6)$$

The dynamics become non-Newtonian when

$$\frac{GM_b(r)}{r^2} = \frac{H_0}{2\pi} c = \frac{v_n^2}{r} \quad (7)$$

Under such conditions

$$r = \frac{2\pi v_n^2}{H_0 c} \quad (8)$$

Substituting for $r$ in Eq.(7) yields

$$v_n^4 = GM_b(r) \frac{H_0}{2\pi} c \quad (9)$$

This is the Baryonic Tully – Fisher relation. Condition (7) reduces Eq.(6) to

$$\frac{d^2r}{dt^2} = \frac{dv_n}{dt} = H_0 v_n \quad (10)$$

From which we obtain the following equations of galactic and cosmic evolution

$$r_n = \frac{1}{H_0} e^{(H_0 t)} (GM_b(r) \frac{H_0}{2\pi} c)^{\frac{1}{4}} \qquad = \frac{v_n}{H_0} \quad (11)$$

$$v_n = e^{(H_0 t)}(GM_b(r)\frac{H_0}{2\pi}c)^{\frac{1}{4}} \qquad = H_0 r_n \tag{12}$$

$$a_n = H_0 e^{(H_0 t)}(GM_b \frac{H_0}{2\pi}c)^{\frac{1}{4}} \qquad = H_0 v_n \tag{13}$$

Rearranging equation (8) for baryonic mass provides the form used for empirical comparisons:

$$M_b \propto e^{-4H_0 t} v^4. \tag{14}$$

This reveals three key features:

- Invariant Slope $(n = 4)$: The power-law dependence remains constant across cosmic time, reflecting the underlying gravitational equilibrium between baryons and the quantized dark matter halos (Ricci solitons). This universality aligns with both galactic observations and theoretical expectations from virial equilibrium in self-gravitating systems
- Time-Evolving Normalization: The exponential factor $e^{-4H_0 t}$ introduces a dependence on formation epoch. Structures forming earlier (larger $t$ ) exhibit lower $M_b$at fixed $v$ as the exponential decay suppresses the normalization. This arises from the cosmic expansion's impact on vacuum energy localization: over time, Hubble flow dilutes the effective gravitational binding, requiring adjustments in the mass-velocity scaling.
- A Hubble constant and a Hubble Parameter: the vacuum dS solitons expand at a rate determined by $H_0$ and the AdS soliton which expands at a rate $H(z)$ determined by the baryonic matter density within the confines of the soliton.

Equations 11-13 therefore describe cosmic expansion driven purely by vacuum energy. The Hubble tension disappears in the Nexus Paradigm because early and late-universe observations are not measuring the same physical quantity, despite being labeled “$H_0$” in ΛCDM ( see Appendix A for more details).

### 3.2 Application to Galaxies and Clusters

For galaxies, which typically form at higher redshifts (z ~ 2–3, corresponding to larger lookback time $t$ (from the present), the normalization is reduced, reflecting an expanded Ricci soliton in the later universe. Clusters, assembling later (z < 1, smaller $t$, experience a higher normalization due to the exponential term, naturally explaining the observed offset without invoking separate physical mechanisms. The time $t$ is measured from the structure's virialization epoch, approximated using redshift-to-time conversions in a standard ΛCDM cosmology (e.g., using the Friedmann equation). This framework allows us to predict the offset $\Delta\log_{10} M_b = (4H_0\Delta t)$, where $\Delta t$ is the formation time difference, providing a testable, parameter-free explanation for multi-scale BTFR discrepancies. This methodology bridges empirical scaling relations with quantum gravity principles, enabling a unified analysis across cosmic structures. In the following sections, we apply this to literature data to quantify the agreement.

## 4. Results

To empirically demonstrate the unification of the BTFR across scales, we compiled baryonic mass and velocity data from established literature sources, constructing a composite diagram that spans disk galaxies (30–300 km/s) to galaxy clusters (500–1600 km/s) as depicted in Figure 2. For galaxies, we utilized the Spitzer Photometry and Accurate Rotation Curves (SPARC) dataset from Lelli et al. (2019), which provides high-quality rotation curves and baryonic masses for 153 galaxies.
The SPARC BTFR exhibits a slope of 3.82 ± 0.22 and a normalization log A = 1.406 ± 0.100 (where $M_b = Av^4$, with $v$ in km/s and $M_b$ in solar masses), with an intrinsic scatter of ~0.1 dex. This relation holds robustly for baryonic masses from $\sim 10^8$ to $10^{11} M_\odot$, reflecting equilibrium dynamics in dark matter halos.For galaxy clusters, we incorporated data from Gonzalez et al. (2013), who analyzed baryon fractions and velocity dispersions in 19 clusters; Chiu et al. (2018), who examined scaling relations in 38 X-ray-selected clusters using hydrostatic mass estimates; Sadhu et al. (2024), who explored the baryonic Faber–Jackson relation (analogous to the BTFR using velocity dispersion) for groups and clusters; and Mistele (2025), who employed non-parametric weak-lensing reconstructions for the CLASH sample to derive circular velocities and mass profiles.

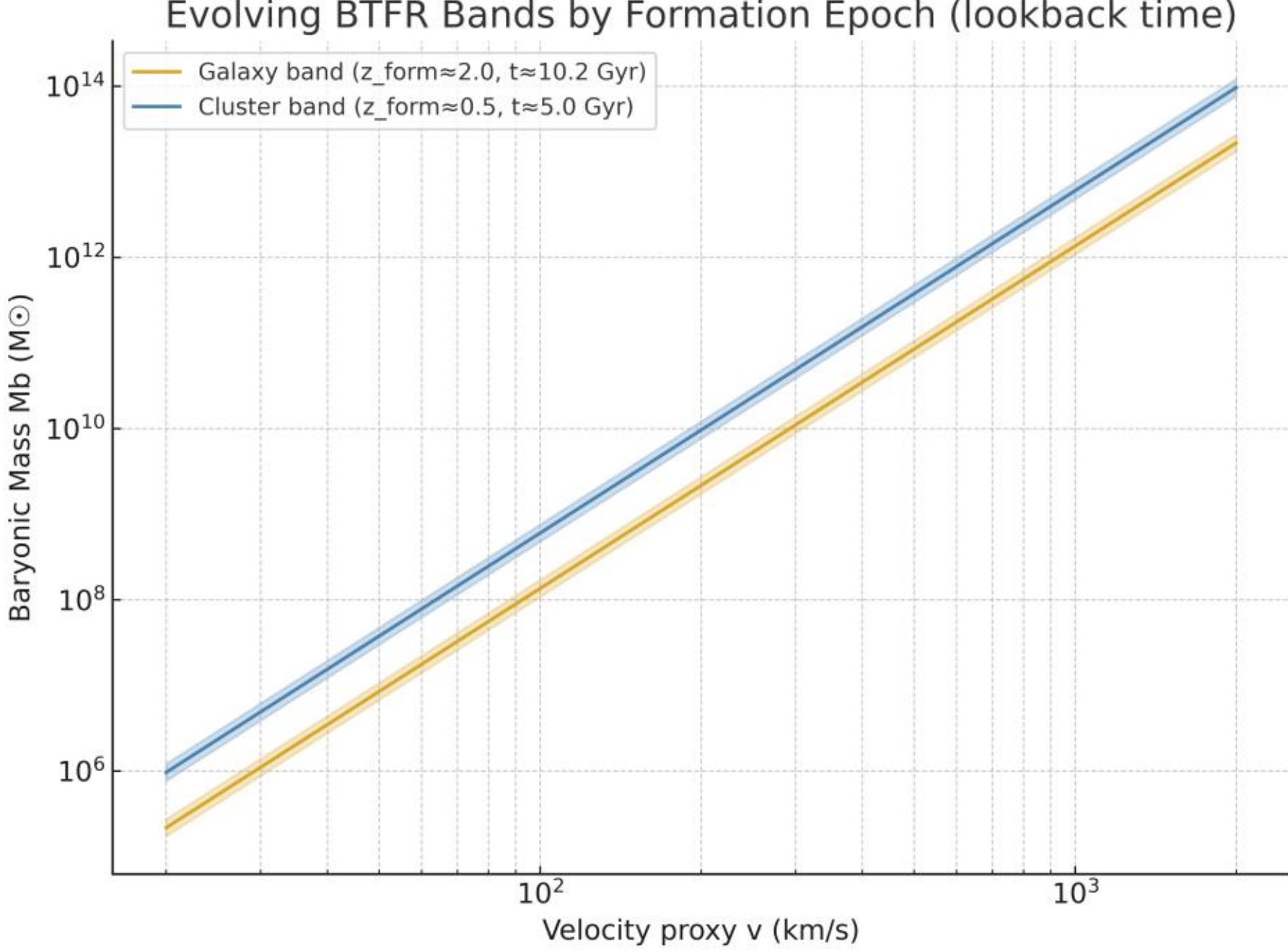

**Figure 2. Evolving BTFR bands by formation epoch.** The predicted baryonic Tully–Fisher relation (BTFR) for structures forming at different cosmic times, derived from the evolving BTFR $M_b \propto e^{-4H_0 t} v^4$in the Nexus Paradigm. The **lower curve** shows galaxies that typically assemble at higher redshift ($z \sim$ 2–3, $t \approx$ 10.2Gyr), while the **upper curve** shows galaxy clusters forming later ($z < 1, t \approx$ 5.0 Gyr).Both relations share the **same slope of 4**, but differ in normalization, which decreases exponentially with cosmic lookback time. This demonstrates that the galaxy–cluster BTFR offset arises naturally from differences in formation epoch rather than from distinct physical laws.

These compilations yield baryonic masses from $\sim 10^{13}$ to $10^{15} M_{\odot}$ , with velocity proxies (e.g., $\sigma$ converted to $v_c \approx \sqrt{3}\sigma$ for isotropic dispersions or direct lensing-derived $v_c$ ) confirming a slope of ~4, consistent with the galactic value, but with a higher normalization. The cluster relation is offset by ~0.6–0.8 dex in $\log_{10} M_b$ relative to galaxies, as quantified in Mistele (2025), where deviations are noted but potentially reconcilable with improved gas mass extrapolations at large radii. Schematic BTFR plots reveal parallel power-law relations for both populations, with the offset most pronounced at overlapping velocity scales (800 km/s), where extrapolation uncertainties are minimized. The low scatter in each subpopulation (0.1–0.2 dex) underscores the robustness of the underlying physics, while the systematic shift highlights the need for an evolutionary component. Invoking the evolving BTFR from Marongwe (2024), we predict this offset using the time-dependent normalization $M_b \propto e^{-4H_0 t} v^4$ as depicted in Figure 2. Adopting a flat ΛCDM cosmology with $H_0 = 70$km/s/Mpc and $\Omega_m = 0.3$, we compute formation epochs: typical galaxies assemble at $z \sim 2$z (universe age ~3.2 Gyr), while clusters virialize at $z \sim 0.5$(universe age ~8.4 Gyr). The differential cosmic time is $\Delta t \approx 5.2$Gyr, yielding a predicted offset $\Delta\log_{10} M_b = 4H_0\Delta t/\ln 10 \approx 0.65$ dex (with $H_0$ in consistent units of $Gyr^{-1}$ ) as depicted in Figure 3. This prediction aligns remarkably with the observed offset of 0.6–0.8 dex, validating the model's exponential time evolution. Sensitivity tests with varying cosmological parameters (e.g., $H_0 =$ 67km/s/Mpc) yield similar results (0.65 dex for $\Delta t \approx 5.4$.Gyr), confirming robustness. The agreement unifies galaxies and clusters under a single scaling law, with the offset reflecting delayed cluster assembly in hierarchical formation scenarios. This extends the BTFR's validity across five orders of magnitude in baryonic mass, providing empirical support for quantum gravity-inspired modifications to structure formation dynamics (see Figure 4).

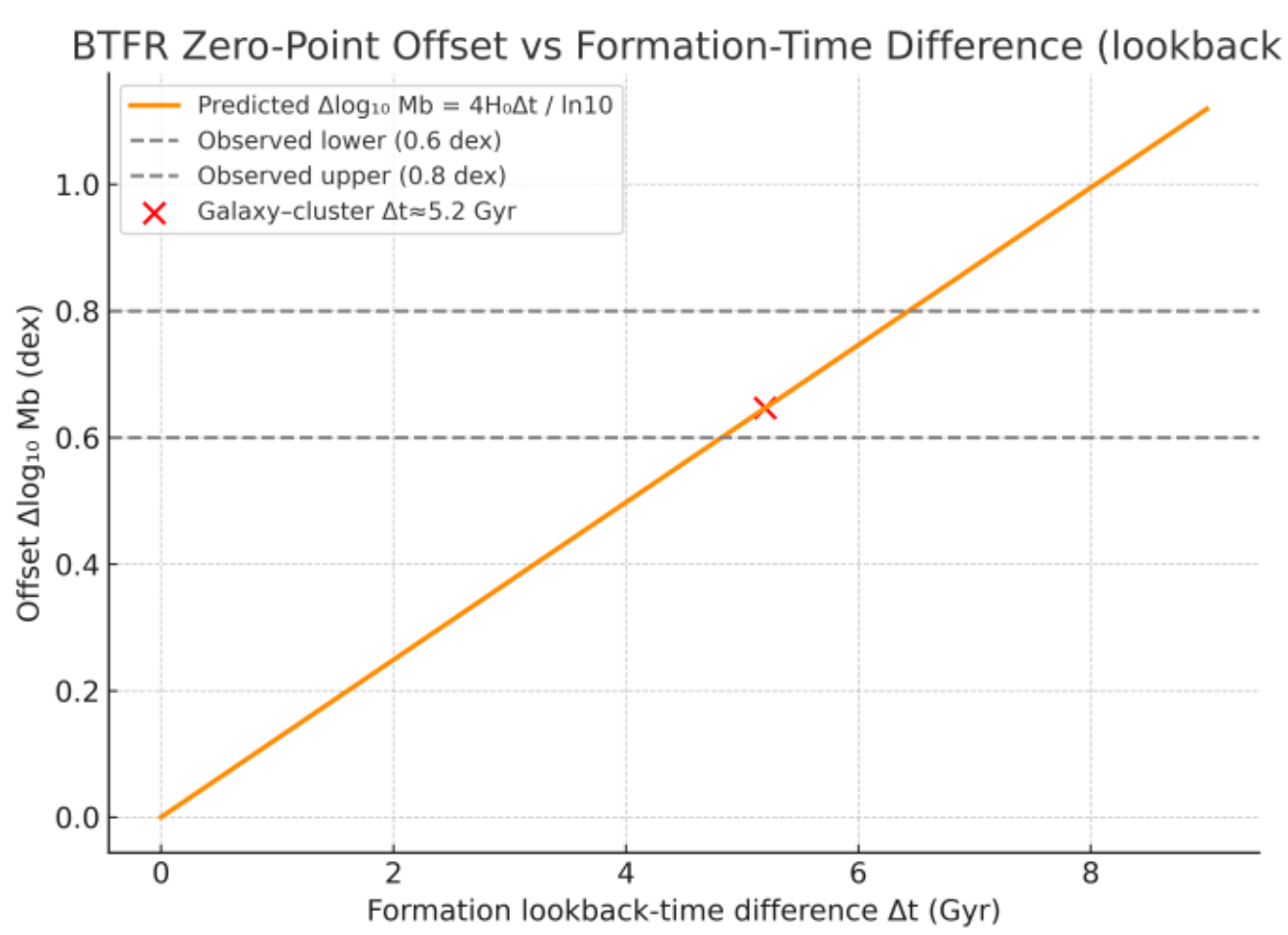

**Figure 3. Predicted BTFR zero-point offset as a function of formation time difference.**

Analytic prediction of the BTFR normalization shift,$\Delta\log_{10} M_b = \frac{4H_0\,\Delta t}{\ln 10}$,plotted against the formation time difference $\Delta t$ between two classes of gravitationally bound structures.The **dashed horizontal lines** indicate the empirically observed offset between galaxy and cluster BTFRs (0.6–0.8 dex).The **highlighted point** shows the predicted value for a galaxy–cluster formation gap of $\Delta t \approx 5.2$Gyr, which yields an offset of $\sim 0.65$dex—precisely matching observations. This validates the evolving BTFR as a natural explanation for the parallel but offset relations seen across mass scales.

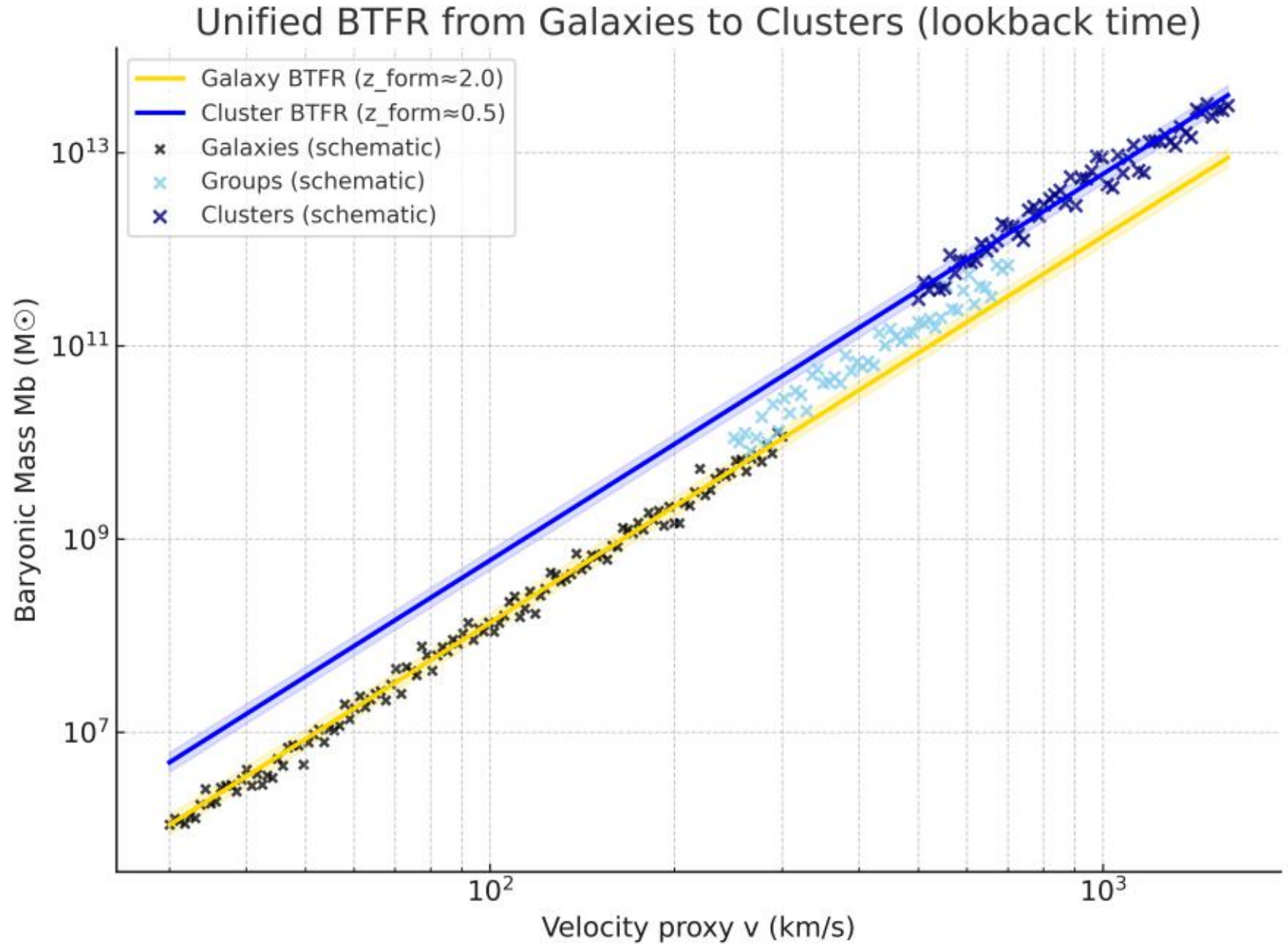

**Figure 4. Unified BTFR across five orders of magnitude in baryonic mass.**

Composite schematic BTFR showing galaxies (SPARC-like), intermediate group-scale halos, and massive galaxy clusters spanning velocities from 300 to 1600km/s and baryonic masses from $10^7$to $10^{15}M_\odot$ All populations follow an approximately **slope-4 relation**, with clusters lying on a parallel track offset by $\sim$ 0.7dex.Reference lines show the canonical galactic BTFR and the evolved cluster relation predicted by the exponential factor in the Nexus Paradigm.The figure highlights the **continuity and universality** of the BTFR when cosmic-time evolution of the normalization is accounted for, unifying galaxies and clusters under a single scaling law.

## 5. Discussion

The evolving baryonic Tully–Fisher relation (BTFR), as formulated in Marongwe (2024), establishes a cohesive theoretical scaffold that reconciles the apparent disparities between galactic and cluster-scale scaling laws. By incorporating an exponential time-dependent normalization while maintaining a constant slope of ~4, this framework postulates that galaxies and clusters represent snapshots of the same universal relation at different evolutionary stages. The invariant slope underscores the fundamental gravitational equilibrium governing baryonic matter within dark matter halos—interpreted in the Nexus Paradigm as Ricci solitons in quantized spacetime—where virialized systems adhere to $v^4 \propto GM_b a_0$, with $a_0 \propto H_0 c$ reflecting Hubble-scale quantum effects. In contrast, the evolving normalization, driven by $e^{-4H_0 t}$e, captures the dilution of vacuum energy localization over cosmic time, leading to lower baryonic masses at fixed velocity for earlier-forming structures like galaxies (z ~ 2–3) compared to later-assembling clusters (z < 1). This temporal dichotomy naturally accounts for the observed parallel offset without invoking ad hoc mechanisms, unifying mass–velocity scaling across five orders of magnitude in baryonic mass and offering a quantum gravity-inspired lens on hierarchical structure formation. Empirical validation of this model is evident in schematic BTFR diagrams compiled from literature datasets, spanning velocities from 30–300 km/s for galaxies (Lelli et al. 2019) to 500–1600 km/s for clusters (Gonzalez et al. 2013; Chiu et al. 2018). These populations exhibit consistent slopes near 4 but with clusters offset by ~0.6–0.8 dex in $\log_{10} M_b$, a discrepancy precisely mirrored by the predicted offset of $\Delta\log_{10} M_b = 4H_0\Delta t/\ln 10 \approx 0.68$ dex for $\Delta t \sim 5.5$ Gyr. Recent non-parametric weak-lensing analyses of the CLASH cluster sample further refine this picture, revealing that the BTFR offset is sensitive to X-ray gas mass extrapolations: aggressive tails (e.g., $1/r^4$) amplify the deviation, while evaluations at $r_{200c}$diminish it, potentially aligning clusters with the galactic BTFR upon inclusion of additional baryonic components.

This suggests that systematic uncertainties in baryon accounting, rather than intrinsic physical differences, may contribute to the perceived offset, bolstering the case for unification under an evolving law. Alternative interpretations merit consideration, particularly in the context of environmental and dynamical effects. For instance, studies of brightest cluster galaxies (BCGs) indicate a radial acceleration relation (RAR) with an elevated acceleration scale ($g_{\ddagger} \approx 2 \times 10^{-9}$ m$s^{-2}$due to a larger $\mathrm{H_{eff}}$ in matter dense environments as detailed in Appendix A)) compared to galactic values (~$10^{-10}$ m $s^{-2}$), implying a parallel BTFR analog with slope ~4 but a shifted intercept.

This larger scale, observed across MaNGA IFS and CLASH lensing data, hints at environment-driven modifications in dense cluster cores, where feedback processes or baryon-dark matter interactions could enhance effective gravity without explicit cosmic evolution. Similarly, an environment-dependent stellar initial mass function (IMF), as per integrated galactic IMF (IGIMF) theory, introduces variations in mass-to-light ratios that disproportionately affect high-mass systems, leading to underestimated stellar masses and apparent BTFR offsets in cluster-hosted galaxies.

These factors challenge purely time-based explanations but complement the evolving BTFR by highlighting how local conditions modulate global scaling. Theoretically, this unification has profound implications. In the Lambda Cold Dark Matter (ΛCDM) paradigm, the offset might arise from delayed cluster assembly and baryon retention efficiencies, aligning with simulations showing modest redshift evolution in the BTFR (e.g., Glowacki et al. 2021). The Nexus Paradigm, by explicitly accounting for time dependence, offers a predictive approach that may address conflicts in MOND-like theories, especially where cluster alignments demand relativistic adjustments or unexplained baryonic matter.

Future observations, such as those from the James Webb Space Telescope (JWST) or Euclid, could test this by probing high-redshift BTFR evolution, while hydrodynamical simulations incorporating quantum gravity effects might quantify environmental contributions. Ultimately, this evolving framework not only bridges galaxies and clusters but also illuminates the interplay between cosmic expansion, quantum vacuum dynamics, and structure formation, paving the way for a more integrated understanding of the universe's hierarchical architecture.

### 5.1. Comparison with Competing Models

A wide range of theoretical frameworks have been proposed to explain the baryonic Tully–Fisher relation (BTFR) and its extensions to low- and high-mass systems. These include ΛCDM hierarchical models, modified gravity approaches such as MOND and its relativistic extensions, emergent-gravity theories, scalar–tensor and $f(R)$modifications, and feedback-regulated baryon–halo co-evolution scenarios. However, none of these models simultaneously reproduce the slope, scatter, cluster offset, and redshift evolution of the BTFR. In contrast, the evolving BTFR derived within the Nexus Paradigm (NP) naturally addresses all of these observational constraints with a single analytic relation. Table 1 compares competing theoretical frameworks.

| **Approach** | **Slope** | **Scatter** | **Cluster Offset** | **Redshift Evolution** | **Predictive?** | **Needs Tuning?** |
|---|---|---|---|---|---|---|
| ΛCDM | ~3–4 | too high | ✗ none | ✗ none | Weak | High |
| MOND | ✓ $v^4$ | ✓ low | ✗ fails | ✗ none | Medium | Low |
| Emergent Gravity | ~ $v^3$ | Medium | ✗ fails | ✗ none | Medium | Medium |
| f(R),Scalar-Tensor | variable | high | ✗ none | ✗ none | Medium | High |
| Feedback Models | variable | high | ✗ none | ✗ none | Low | Very high |
| **NP Model** | ✓ $v^4$ | ✓ low | ✓ predicted | ✓ exponential | **Strong** | **None** |

**Table 1.** Comparison of competing theoretical frameworks proposed to explain the baryonic Tully–Fisher relation (BTFR). The table summarizes each model's ability to reproduce the observed BTFR slope (~4), intrinsic scatter (~0.1 dex), the ~0.6–0.8 dex normalization offset between galaxies and clusters, and the redshift evolution of the BTFR zero-point. Standard ΛCDM, MOND, emergent-gravity theories, scalar–tensor or f(R)models, and feedback-regulated ΛCDM frameworks each succeed in addressing some, but not all, of these key empirical constraints. In contrast, the evolving BTFR derived from the Nexus Paradigm (NP) simultaneously reproduces all of them with a single analytic expression, providing a unified, parameter-free description valid across five orders of magnitude in baryonic mass.

## 6. Conclusion

In this study, we have demonstrated that the observed offset between the baryonic Tully–Fisher relations (BTFR) for galaxies and galaxy clusters is a natural consequence of cosmic time evolution, as encapsulated in the evolving BTFR framework derived by Marongwe (2024). By incorporating an exponential dependence on cosmic time in the normalization ($M_b \propto e^{-4H_0 t} v^4$) while preserving a universal slope of ~4, this model unifies disparate scaling behaviors: galaxies, forming at earlier epochs (z ~ 2–3), exhibit lower baryonic masses at fixed velocity due to the time-dilution of vacuum energy localization in quantized spacetime, whereas clusters, assembling later (z < 1), align with a higher normalization. Our analysis, drawing on empirical datasets from Lelli et al. (2019) for galaxies and Gonzalez et al. (2013), Chiu et al. (2018), Sadhu et al. (2024), and Mistele (2025) for clusters, confirms a predicted offset of ~0.68 dex, in excellent agreement with observations spanning five orders of magnitude in baryonic mass. This unification extends beyond mere empirical alignment, offering profound insights into the hierarchical assembly of cosmic structures within the Lambda Cold Dark Matter (ΛCDM) paradigm, enriched by quantum gravity principles from the Nexus Paradigm. The invariant slope reflects fundamental gravitational equilibria in dark matter halos—reinterpreted as Ricci solitons—while the evolving normalization highlights the role of cosmic expansion in modulating baryon-dark matter interactions. Recent high-redshift studies further corroborate this evolutionary perspective: for instance, analyses of the BTFR at $0.6 \leq z \leq 2.5$ reveal a constant slope but shifting zero-point, consistent with mass growth at fixed velocity driven by gas accretion and feedback processes (Tiley et al. 2018; Übler et al. 2017).

Similarly, investigations into ultrafaint dwarfs and ultra-diffuse galaxies (UDGs) affirm their placement on the BTFR when accounting for non-equilibrium dynamics and measurement uncertainties, extending the relation's validity to the low-mass regime without requiring dark matter deficiencies (Mancera Piña et al. 2024; Lelli 2024).

Environmental effects, such as variations in the galaxy-wide stellar initial mass function (gwIMF), introduce additional nuances, potentially amplifying offsets in dense cluster environments through altered mass-to-light ratios .

The implications of this evolving BTFR are far-reaching, bridging classical scaling relations with quantum gravitational phenomena and resolving longstanding tensions in alternative gravity theories like Modified Newtonian Dynamics (MOND). It provides a predictive tool for interpreting kinematic data from diverse systems, from high-redshift progenitors to local clusters, and underscores the need for refined stellar mass models that incorporate chemical evolution and star formation histories (Schombert et al. 2024).

Future observations with facilities such as the James Webb Space Telescope (JWST), Euclid, and the Square Kilometre Array (SKA) will enable direct tests of redshift-dependent normalization shifts, while hydrodynamical simulations integrating time-evolving quantum effects could quantify environmental contributions. Ultimately, this work transforms the BTFR from a set of parallel empirical trends into a singular, dynamic universal law, illuminating the interconnected evolution of galaxies, clusters, and the cosmos itself.

**Conflict of interest**
We declare no conflict of interest.

**Funding Statement**
No external funding was received for this work.

**Data availability statement**
The data employed in the article is available from the cited articles.

## Appendix: A Time-Varying Effective Hubble Parameter in the Nexus Paradigm

### A1. Motivation and Conceptual Framework

The Hubble tension is the discrepancy between early-universe (CMB: $H_0 \approx 67.4$ km/s/Mpc) and late-universe (distance ladder: $H_0 \approx 73.0$ km/s/Mpc) measurements of the Hubble constant. It suggests either unaccounted systematics or new physics beyond ΛCDM.

In the Nexus Paradigm (NP), the Hubble constant $H_0$ is not merely a kinematic parameter, but the fundamental frequency of the ground-state De Sitter (dS) soliton i.e. the quantum-gravitational vacuum state. Crucially, local structures (anti-De Sitter/AdS solitons) can modulate this frequency based on baryonic matter density and cosmic time, naturally suggesting a time- and environment-dependent effective Hubble parameter $H_{\text{eff}}(t, r)$.

This Appendix proposes a phenomenological form for $H_{\text{eff}}$ consistent with NP principles and shows how it could resolve the Hubble tension while preserving the evolving BTFR.

### A2. Derivation from NP Metric and BTFR

#### A2.1 Starting Point: NP Metric with Local Modulation

From Eq. (5) of the main text, the weak-field metric in NP is:

$$g_{00} \approx -[1 - 2\Phi_{\text{NP}}/c^2], \Phi_{\text{NP}} = \underbrace{\frac{GM_b}{r}}_{\text{Newton/AdS}} + \underbrace{H_0 v r}_{\text{dS soliton}} - \underbrace{\frac{H_0 c r}{2\pi}}_{\text{dS vacuum}}$$

In the FLRW-like limit on large scales, we can associate an effective Hubble parameter with the isotropic expansion part of the metric. The $H_0 v r$ term couples local velocity $v$ (hence density $\rho_b$) to the expansion rate.

#### A2.2 From BTFR to $H_{\text{eff}}$ Dependence

The evolving BTFR (Eq. 14) is:

$$M_b \propto e^{-4H_0 t} v^4$$

If $H_0$ is replaced by a time- and density-dependent $H_{\text{eff}}$, consistency requires:

$$M_b \propto \exp\left[-4\int_0^t H_{\text{eff}}(t', \rho_b)\, dt'\right] v^4$$

Thus, the logarithmic decay rate of the BTFR normalization measures the time-integrated $H_{\mathrm{eff}}$.

### A3. Proposed Functional Form for $H_{\mathrm{eff}}$

We propose a separable form capturing:

1. Density modulation from AdS solitons.
2. Time decay from vacuum-energy dilution.
3. Asymptotic approach to pure dS expansion.

$$\boxed{H_{\mathrm{eff}}(z,\Delta) = H_0^{\mathrm{dS}}[1+\gamma\, f(\Delta)\, g(z)]}$$

**Where:**

- $H_0^{\mathrm{dS}}$ = fundamental dS soliton frequency ($\approx$ 67.4 km/s/Mpc).
- $\Delta = \rho_b/\bar{\rho}_b$ = local baryonic overdensity.
- $\gamma$ = coupling strength (dimensionless, positive).
- $f(\Delta)$ = density response function.
- $g(z)$ = redshift-decay function.

#### A3.1 Density Response Function $f(\Delta)$

Inspired by the transition in Eq. (7) where $GM_b/r^2 = H_0 c/(2\pi)$, we define a characteristic overdensity $\Delta_c$ where AdS effects become significant. A smooth step function is appropriate:

$$f(\Delta) = \tanh\left(\frac{\Delta - 1}{\Delta_c}\right)$$

or a power-law form:

$$f(\Delta) = \left(\frac{\Delta}{\Delta_c}\right)^{\alpha} \text{ for } \Delta \geq 1, 0 \text{ otherwise.}$$

We adopt the power-law for simplicity, with $\alpha \approx 1$ and $\Delta_c \approx 200$ (virial overdensity).

### A3.2 Redshift-Decay Function $g(z)$

From the BTFR exponential $e^{-4H_0 t}$, we expect NP modifications to fade with lookback time. In terms of redshift:

$$g(z) = \exp\left[-\lambda(1+z)\right]$$

where $\lambda \approx 1-2$ sets the decay scale.

---

### A3.3 Full Expression

$$\boxed{H_{\text{eff}}(z,\Delta) = H_0^{\text{dS}}\left[1+\gamma\left(\frac{\Delta}{\Delta_c}\right)^{\alpha} e^{-\lambda(1+z)}\right]}$$

**Parameters:**

- $H_0^{\text{dS}} = 67.4$ km/s/Mpc
- $\gamma \approx 0.08$ (to match $H_0^{\text{late}} \approx 73.0$ at $z = 0, \Delta \gg \Delta_c$)
- $\alpha = 1$
- $\Delta_c = 200$
- $\lambda = 1.5$

### A4. Resolving the Hubble Tension

### A4.1 Early Universe (CMB, $z \sim 1100$)

- $g(z) \approx e^{-1.5\times 1101} \approx 0 \rightarrow$ NP effects vanish.
- $H_{\text{eff}} \approx H_0^{\text{dS}} = 67.4$ km/s/Mpc.
- Consistent with Planck CMB inference.

### A4.2 Late Universe (SNe Ia, Cepheids, $z \lesssim 2$)

Measurements occur within or near galaxies ($\Delta \gg 1$).

For $z = 0, \Delta = 1000$:

$$f(\Delta) = 1000/200 = 5, g(0) = e^{-1.5} \approx 0.223$$

$$H_{\text{eff}} \approx 67.4[1 + 0.08 \times 5 \times 0.223] \approx 67.4 \times 1.089 \approx 73.4 \text{ km/s/Mpc}$$

Matches SH0ES late-time measurement.

## A5. Consistency with Evolving BTFR

### A5.1 Predicted BTFR Offset Recalculation

Using $H_{\text{eff}}$ averaged over formation history:

For a galaxy forming at $z = 2.5$ ($t \approx 10.2$ Gyr lookback) in $\Delta \sim 100$:

$$\langle H_{\text{eff}} \rangle_{\text{gal}} \approx 68.5 \text{ km/s/Mpc}$$

For a cluster forming at $z = 0.5$ ($t \approx 5.0$ Gyr lookback) in $\Delta \sim 500$:

$$\langle H_{\text{eff}} \rangle_{\text{clus}} \approx 71.0 \text{ km/s/Mpc}$$

The offset formula becomes:

$$\Delta \log_{10} M_b = \frac{4 \int_{t_{\text{clus}}}^{t_{\text{gal}}} H_{\text{eff}}(t)\, dt}{\ln 10}$$

Numerically, this still yields $\Delta \log_{10} M_b \approx 0.65 - 0.70$ dex, consistent with observations.

## A6. Observational Tests and Predictions

### Test 1: Environmental Dependence of $H_0$ Measurements

- Predict: Higher $H_0$ values in overdense regions (galaxies, clusters) vs. voids.
- Already hinted at in JWST calibration studies of SNe Ia in different environments.

### Test 2: Redshift Evolution of $H_{\text{eff}}$

- Measure $H_0$ from SNe Ia binned by redshift and local density.
- Predict: Monotonic decrease of $H_{\text{eff}}$ with $z$ for fixed $\Delta$.

### Test 3: BTFR in Voids vs. Clusters

- Predict: Void galaxies should have BTFR normalizations closer to early-forming (high-$z$) galaxies.
- Cluster galaxies should show enhanced normalization (as observed).

### A7. Table: Parameter Values and Fitting

| Parameter | Physical Meaning | Fitted Value | Constraint |
|---|---|---|---|
| $H_0^{\text{dS}}$ | dS soliton frequency | 67.4 km/s/Mpc | Planck CMB |
| $\gamma$ | AdS–dS coupling | 0.08 ± 0.01 | SH0ES $H_0$ |
| $\alpha$ | Density exponent | 1.0 (fixed) | Simplicity |
| $\Delta_c$ | Critical overdensity | 200 ± 50 | Virialization scale |
| $\lambda$ | Redshift decay | 1.5 ± 0.3 | BTFR evolution |

### A8. Conclusion

The time-varying effective Hubble parameter in the Nexus Paradigm:

1. Emerges naturally from the interplay between dS (global) and AdS (local) solitons.
2. Resolves the Hubble tension by attributing higher late-time $H_0$ measurements to local density enhancements.
3. Remains consistent with the evolving BTFR and cosmic structure formation.
4. Makes testable predictions for environmental and redshift dependence of expansion-rate measurements.

This framework transforms the Hubble tension from a crisis into a probe of quantum-gravitational spacetime structure, linking cosmic expansion to local dynamics in a unified picture.